\documentclass[preprint,3p,10pt]{elsarticle} 

\usepackage{graphicx,color}
\usepackage[normalem]{ulem} 
\usepackage{url}

\DeclareUrlCommand\ULurl{%
  \renewcommand\UrlLeft{\uline\bgroup}%
  \renewcommand\UrlRight{\egroup}}
\usepackage[utf8]{inputenc}
\usepackage[T1]{fontenc}
\usepackage{amsmath,amssymb,amsfonts,textcomp}
\usepackage{calc}
\usepackage{hyperref}
\usepackage{array}
\usepackage{booktabs}
\usepackage{caption}
\usepackage{helvet}

\usepackage{amsmath}
\usepackage{nccmath}
\usepackage{hyperref}
\usepackage{wrapfig} 
\usepackage{array}
\newcolumntype{x}[1]{>{\centering\arraybackslash\hspace{0pt}}p{#1}}
\usepackage{textcomp}
\usepackage{afterpage}
\usepackage{enumitem}
\usepackage[percent]{overpic}
\usepackage{pgfplots} 
\usepackage{tikz}
\usepgfplotslibrary{groupplots}
\usetikzlibrary{external,matrix,arrows,positioning}
\usetikzlibrary{
        pgfplots.external,
        pgfplots.groupplots,
    }
\pgfplotsset{compat=1.3}
\pgfplotsset{compat=newest} 
\pgfplotsset{plot coordinates/math parser=false}
\pgfplotsset{select coords between index/.style 2 args={
    x filter/.code={
        \ifnum\coordindex<#1\def\pgfmathresult{}\fi
        \ifnum\coordindex>#2\def\pgfmathresult{}\fi
    }
}}
\definecolor{color1}{RGB}{228,26,28}
\definecolor{color2}{RGB}{55,126,184}
\definecolor{color3}{RGB}{77,175,74}
\definecolor{color4}{RGB}{152,78,163}
\definecolor{color5}{RGB}{255,127,0}
\definecolor{color6}{RGB}{255,255,51}
\definecolor{redVLE}{RGB}{213,62,79}
\definecolor{blueVLE}{RGB}{50,136,189}

\makeatletter
\def\ps@pprintTitle{%
 \let\@oddhead\@empty
 \let\@evenhead\@empty
 \def\@oddfoot{}%
 \let\@evenfoot\@oddfoot}
\makeatother

\begin{document}

\begin{frontmatter}

\title{\textbf{Experimental and Numerical Investigation of Phase Separation\\ due to Multi-Component Mixing at High-Pressure Conditions\tnoteref{t1}}}


\tnotetext[t1]{Accepted manuscript, ILASS 2017 in Valencia, Spain}

\author[UniBw]{C.~Traxinger\corref{cor1}}
\cortext[cor1]{Corresponding author: christoph.traxinger@unibw.de}
\author[UniBw]{H.~Müller\corref{cor2}}
\cortext[cor2]{Now at: MTU Aero Engines AG, hagen.mueller@mtu.de}
\author[UniBw]{M.~Pfitzner}
\address[UniBw]{Institute for Thermodynamics, Bundeswehr University Munich, Germany}
\author[ITLR]{S.~Baab}
\author[ITLR]{G.~Lamanna}
\author[ITLR]{B.~Weigand}
\address[ITLR]{Institute of Aerospace Thermodynamics, University of Stuttgart, Germany}
\author[TUM]{\\J.~Matheis}
\author[TUM]{C.~Stemmer}
\author[TUM]{N.~A.~Adams}
\address[TUM]{Chair of Aerodynamics and Fluid Mechanics, Technical University of Munich, Germany}
\author[TUDelft]{S.~Hickel}
\address[TUDelft]{Faculty of Aerospace Engineering, Technische Universiteit Delft, The Netherlands\\\vspace*{-16pt}}

\begin{abstract}
Experiments and numerical simulations were carried out in order to contribute to a better understanding and prediction of high-pressure injection into a gaseous environment. Specifically, the focus was put on the phase separation processes of an initially supercritical fluid due to the interaction with its surrounding. N-hexane was injected into a chamber filled with pure nitrogen at 5 MPa and 293 K and three different test cases were selected such that they cover regimes in which the thermodynamic non-idealities, in particular the effects that stem from the potential phase separation, are significant. Simultaneous shadowgraphy and elastic light scattering experiments were conducted to capture both the flow structure as well as the phase separation. In addition, large-eddy simulations with a vapor-liquid equilibrium model were performed. Both experimental and numerical results show phase formation for the cases, where the a-priori calculation predicts two-phase flow. Moreover, qualitative characteristics of the formation process agree well between experiments and numerical simulations and the transition behaviour from a dense-gas to a spray-like jet was captured by~both.
\end{abstract}

\end{frontmatter}

\section{Introduction}
Injection into a high-pressure gaseous environment is a crucial process within energy conversion machines. Nowadays, many fluid flow devices are operated at pressures that exceed the critical pressure $p_c$ of the involved pure fluids. The increase in operating pressure in aircraft and car engines mainly stems from the demand for higher engine efficiency and reduced CO$_2$ emissions. The main reason for rising the chamber pressure in liquid rocket engines (LREs) is the proportionality between operating pressure and specific impulse \cite{chehroudi2012a}. Typically, the operating pressure in LREs is supercritical with respect to both fuel and oxidizer ($p > p_c$), whereas the injection temperature may be sub- or supercritical, corresponding to liquid-like or gas-like states. At supercritical pressure, the fluid properties, such as density, enthalpy and viscosity, are highly non-linear functions of temperature and pressure. Furthermore, phase separation due to non-linear interaction of the different components may occur. The phenomenon of phase separation due to mixing at high pressures is well-known in process engineering. Remarkably, up to now, high-pressure fuel injection into a gaseous environment is not completely understood and no commonly accepted theoretical approach~exists.\\
Within the past 20 years, many research groups have focused on understanding the behaviour of jets at high pressures using experimental and numerical methods. Chehroudi~et~al.~\cite{chehroudi1999a} injected cryogenic nitrogen into gaseous nitrogen at sub- and supercritical pressures. Based on shadowgraphy visualizations, they observed classical two-phase phenomena at subcritical pressure indicated by very fine ligaments and droplets being ejected from the jet. As the pressure exceeds the critical value, surface tension effects diminish and the enthalpy of vaporization disappears. As a consequence, droplets were no longer detected and finger-like structures were observed on the jet surface. Similar phenomena occur in multi-component mixtures. Mayer~et~al.~\cite{mayer1998a} investigated coaxial LN$_2$/GHe injection into GHe and reported a drastic change of the interfacial structure depending on the chamber pressure. In the low pressure case, a liquid spray with droplets was formed, whereas the breakup seemed to transit into gas-like turbulent mixing for pressures significantly beyond $p_c$. Another interesting situation for two-phase disintegration was experimentally investigated by Roy~et~al.~\cite{roy2013a}. Here, an initially supercritical fluid was injected into a supercritical pressure environment. It was found that, for sufficiently low ambient temperatures, the jet undergoes phase separation leading to the formation of droplets and ligaments in the jet. As stated above, this mainly stems from the interaction between the injectant and the surrounding gas that, in turn, demands for an accurate thermodynamic framework in order to investigate high-pressure injection numerically. Popular numerical test cases are the jets investigated by Mayer~et~al.~\cite{mayer2003a}, where cold (cryogenic) nitrogen is injected into a warm nitrogen environment at supercritical pressure and thus phase separation is not likely to occur. These jets were numerically investigated by different groups, e.g., Zong~et~al.~\cite{zong2004a}, Schmitt~et~al.~\cite{schmitt2009a} and Müller~et~al.~\cite{muller2016b}, using the commonly accepted dense-gas approach. A multi-component benchmark case is the Spray-A of the Engine Combustion Network (\url{www.sandia.gov/ECN}), where cold n-dodecane is injected into a warm nitrogen atmosphere at a pressure of 6 MPa, which exceeds the critical pressure of both components. Different modelling approaches, e.g., Lagrangian particle tracking in Wehrfitz~et~al.~\cite{wehrfritz2013a}, a single-phase dense gas approach in Lacaze~et~al.~\cite{lacaze2015a} and a two-phase approach based on a vapor-liquid equilibrium model in Matheis~and~Hickel~\cite{matheis2016a}, have been used to simulate this test case. Up to now, none of these approaches is commonly accepted because of a common disagreement on the actual state of the fluid or the mixture. This is partially due to the lack of experimental data and well-defined boundary conditions. The latter are highly important as flow phenomena may be very sensitive to thermodynamic (mixture) states.\\
It is the objective of this study to gain a better understanding of high-pressure injection and mixing. Experiments and numerical simulations were conducted, in which accurate synchronization of the thermodynamic boundary conditions was assured. For these investigations, a multi-component system consisting of n-hexane and nitrogen was chosen and a systematic study was carried out at supercritical pressure with respect to the pure components value. The injection temperatures of n-hexane were chosen such that they cover regimes in which the thermodynamic non-idealities, in particular, the effects that stem from the potential phase separation, are significant. The objective of this paper is to investigate whether or not a clear transition from a dense-gas to a spray-like jet with droplets can be observed. The study shows that the large-eddy simulation (LES) in combination with a vapor-liquid equilibrium (VLE) model \cite{matheis2016a} is able to predict the experimental results very well.

\section{Experimental facility and injection system}

\begin{figure}[h]
    \vspace{0pt}
    \centering
    \includegraphics[width=0.70\textwidth]{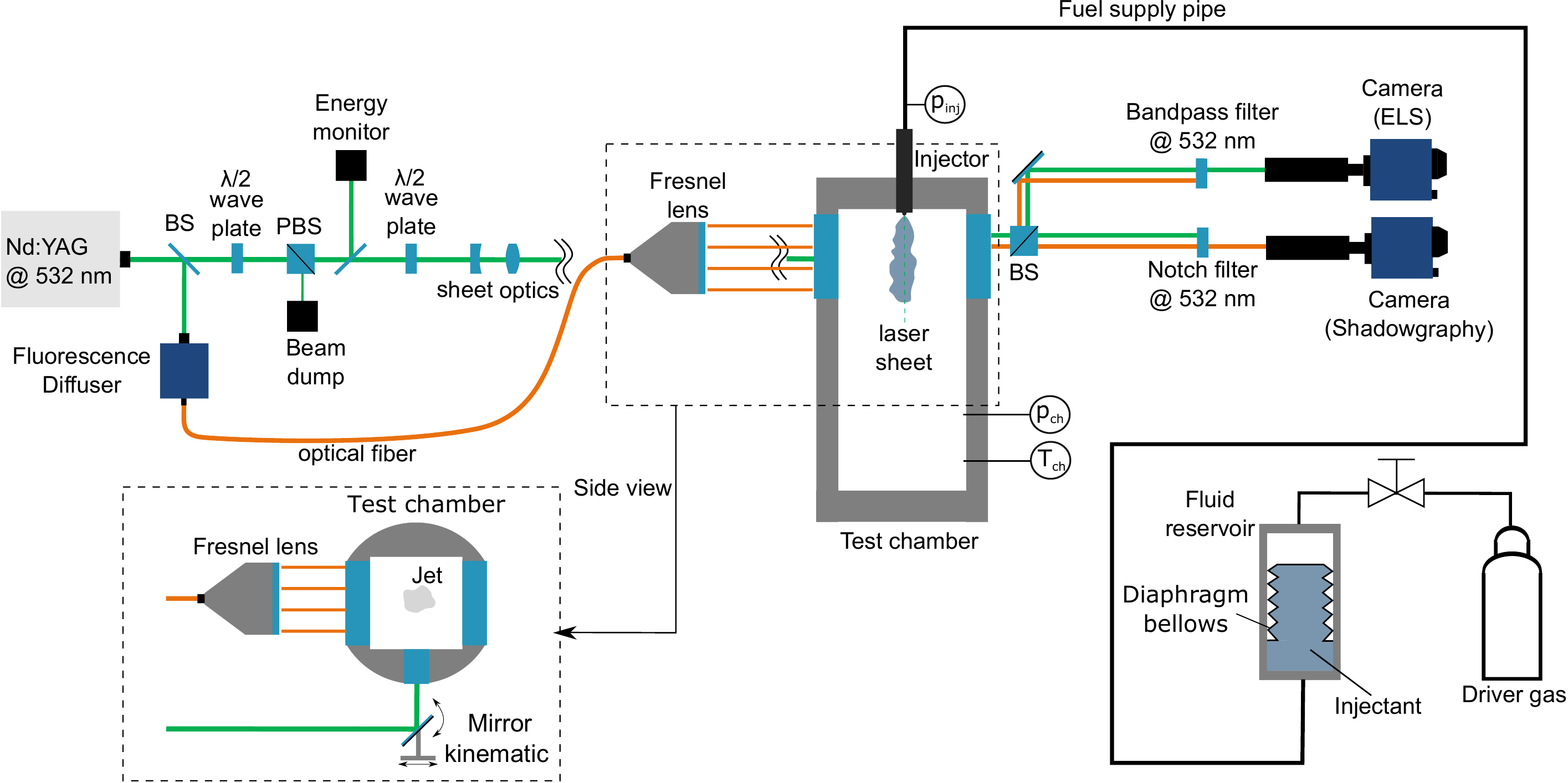}
       \centering
    \caption{Experimental facility including optical arrangement for simultaneous shadowgraphy/ELS measurements.}
    \vspace{-12pt}
    \label{fig:exp_setup}
\end{figure}

The injection experiments of n-hexane into nitrogen ($\geq$ 99.9990\% purity) have been conducted at the ITLR (University of Stuttgart) using the experimental and optical setup that is sketched in Fig.~\ref{fig:exp_setup}. The facility consists of a cylindrical constant-volume chamber ($V\approx 4\cdot 10^{-3}$ m\textsuperscript{3}) that is designed for injections into non-heated ambient gas with a maximum pressure of up to 60~bar. The chamber pressure $p_{ch}$ was measured with a piezoresistive sensor (Keller PA-21Y, 0.25\% uncertainty), while a resistance thermometer provided the chamber temperature within $\pm 1$~K. Three quartz windows (two at the side walls, one at the bottom) enabled optical access into the chamber. A magnetic-valve common-rail injector (distributed by Robert Bosch AG) is mounted to the chamber such that the center axes of the injector nozzle and the cylinder coincide. The injector nozzle is custom-made and features a single straight-hole of diameter $D=0.236$ mm and length $L=0.8$ mm (i.e. $L/D\approx 3.4$). It is important to note that the focus of this study was put on jets with low expansion ratios $p_{inj}/p_{ch}$, which is contrary to conventional Diesel engine applications operating at injection pressures of typically several hundred bar. Consequently, we made adjustments to the needle/spring configuration in order to provide proper needle lift also for comparably low $p_{inj}$ of <~60~bar (as required here). N-hexane was thoroughly degasified prior to the campaign and stored in a fluid reservoir that is separated into two sections using a diaphragm bellows. The section that contained the n-hexane was directly connected to the fuel pipe of the injector. The exterior section was filled with a driver gas to establish the injection pressure, which was also measured with a piezoresistive sensor (Keller PA-23, 0.2\% uncertainty). Using this configuration avoids contamination of the injectant with dissolved driver gas, which is of particular importance for near-critical fluid investigations. The injection temperature $T_{inj}$ was controlled by two heater cartridges with an uncertainty of $\pm 2$ K as detailed in Baab~et~al.~\cite{baab2016a}.

\section{Optical techniques and data processing}

We used parallel-light direct shadowgraphy in combination with planar 90 degree-elastic light scattering (ELS) for the experimental jet analysis. In several recent studies, this choice has been shown suitable to assess both the geometric jet topology as well as the occurrence of two-phase regions embedded within the dense core region \cite{lamanna2012a,baab2014a}. Specifically, scattered signal intensities from a fluid jet illuminated by a light source are very sensitive to the size of present scattering particles. Thus, signal magnitudes vary by several orders of magnitude among single-phase flow (molecular Rayleigh scattering) and two-phase flows with embedded -- potentially very small -- droplets in the particle Rayleigh or Mie regime, Miles~et~al.~\cite{miles2001a}. Proper acquisition and interpretation of the scattered light portion, therefore, allows for a qualitative characterization of multi-phase regions within a fluid flow. 

The used optical arrangement resembles the one used in Baab~et~al.~\cite{baab2014a} in the sense that a single frequency-doubled pulse from an Nd:YAG laser (Continuum Powerlite DLS 8010, $\lambda = 532$ nm) provided illumination for both the shadowgraphy and ELS measurement. It is important to note that this assures instantaneous and simultaneous acquisition of the experimental data. Moreover, stereoscopic image correction of shadowgram and ELS image enables accurate assignment of two-phase phenomena to the local jet region and specific flow features revealed in the shadowgram (e.g. jet core or mixing layer). Both images were captured with a LaVision sCMOS camera (2560 x 2160 pixels, 16-bit dynamic range or 0-65535 digital counts) and a long-distance microscope (Infinity K2 DistaMax) leading to a resolution in the object plane of around 9 \textmu m/pixel. For the ELS measurement, a laser sheet was formed by serial arrangement of a plano-concave and a plano-convex cylindrical lens with focal lengths of -200 and 500 mm, respectively. The laser sheet was reflected into the chamber through the bottom window (see side view in Fig.~\ref{fig:exp_setup}), where the sheet precisely aligned with the nozzle center axis. The excitation intensity could be continuously varied using a $\lambda/2$ wave plate in combination with a polarizing beam splitter cube and was monitored with a pyroelectric energy head. Shadowgraphic illumination was attained by guiding a fraction of the laser pulse into a fluorescence diffuser connected to a Fresnel lens. As the diffuser red-shifts the laser light to a spectrum of approx. 574~-~580~nm, the shadowgraphy and ELS signals could be captured independently by spectral filtering. Specifically, a 532 nm-notch filter removed scattered laser light in the shadowgram, whereas a 532 nm-narrow band-pass filter eliminated the diffuser light in the ELS image. Dark-frame subtraction reduced the noise level in the ELS image below 10 counts. We additionally rescaled the noise-corrected camera intensity $I_{cam}$ (in counts) with the measured excitation intensity $I_0$ according to $\tilde{I} = I_{cam} / I_0$ (counts/mJ) to account for inevitable shot-to-shot variability of the laser. Furthermore, ELS experiments at different magnitudes of $I_0$ remain comparable (in assumption of a linear scattering response). Note, once again, that scattering intensities are subject to strong variations depending on the two-phase jet properties (e.g. droplet sizes and number densities). Consequently, $I_0$ was therefore adjusted to assure proper sensor saturation for all experimental conditions.

\section{Numerical method}

In order to investigate high-pressure injection, large-eddy simulations have been carried out with two different solvers. These solvers have been extensively used for supercritical injection simulations of both pure as well as binary jets, see e.g., \cite{muller2016b,matheis2016a,muller2016a}, whereby one of those solvers is the density-based INCA (\url{www.inca-cfd.com}) and the other is a pressure-based version of OpenFOAM (\url{www.openfoam.org}). For the reason of clarity and with thermodynamic aspects being the main objective -- not the cross-comparison of LES codes -- only one set of LES results (INCA) is presented in this paper. Identical thermodynamic models are implemented in both solvers and are discussed in the following.\\

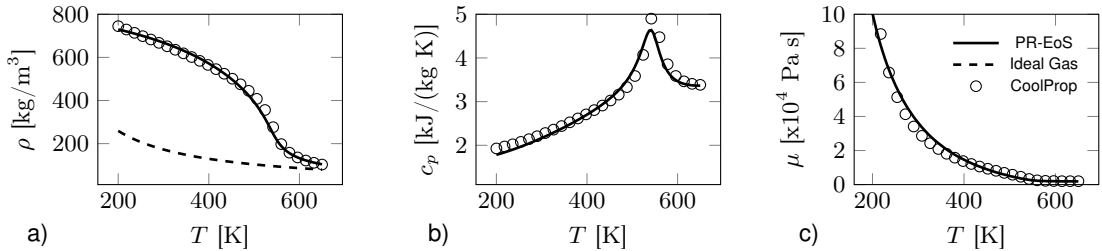
\begin{figure}[t]
\begin{center}
\tikzsetnextfilename{validationPR}
\begin{tikzpicture}
	\begin{groupplot}[group style = {group name = plots,group size = 3 by 1, horizontal sep = 50pt}, width = 0.3\textwidth, height = 0.24\textwidth]
		\nextgroupplot
		[
		ymin = 10,
		ymax = 800,
		ylabel = {$\rho \textrm{ [kg/m$^3$]}$},
		xlabel = {$T \textrm{ [K]}$}
		]
			\addplot [color=black,line width=1.0pt] table [x=T,y=rhoPR]{./data/val_Hexane_p_5MPa.dat};\label{plots:plot1}
			\addplot [color=black,line width=1.0pt,dashed] table [x=T,y=rhoIdGas]{./data/val_Hexane_p_5MPa.dat};\label{plots:plot2}
			\addplot [color=black,each nth point={4},mark=o,only marks,mark options={fill=white}] table [x=T,y=rhoCoolProp]{./data/val_Hexane_p_5MPa.dat};\label{plots:plot3}
			\coordinate (top) at (rel axis cs:0,1);
		\nextgroupplot
		[
		ymin = 1.1,
		ymax = 5,
		ylabel = {$c_p \textrm{ [kJ/(kg K)]}$},
		y filter/.code={\pgfmathparse{#1*1e-3}\pgfmathresult},
		xlabel = {$T \textrm{ [K]}$}
		]
			\addplot [color=black,line width=1.0pt] table [x=T,y=cpPR]{./data/val_Hexane_p_5MPa.dat};
			\addplot [color=black,each nth point={4},mark=o,only marks,mark options={fill=white}] table [x=T,y=cpCoolProp]{./data/val_Hexane_p_5MPa.dat};
		\nextgroupplot
		[
		ymin = 0,
		ymax = 10,
		ylabel = {$\mu \textrm{ [x10$^4$ Pa\;s]}$},
		y filter/.code={\pgfmathparse{#1*1e4}\pgfmathresult},
		xlabel = {$T \textrm{ [K]}$}
		]
			\addplot [color=black,line width=1.0pt] table [x=T,y=muPR]{./data/val_Hexane_p_5MPa.dat};
			\addplot [color=black,each nth point={4},mark=o,only marks,mark options={fill=white}] table [x=T,y=muCoolProp]{./data/val_Hexane_p_5MPa.dat};
			\coordinate (bot) at (rel axis cs:1,0);
	\end{groupplot}	
\node[below = 0.4cm of plots c1r1.south,xshift=-2.4cm] {a)};
\node[below = 0.4cm of plots c2r1.south,xshift=-2.1cm] {b)};
\node[below = 0.4cm of plots c3r1.south,xshift=-2.2cm] {c)};
\path (top|-current bounding box.north)--
      coordinate(legendpos)
      (bot|-current bounding box.north);
\matrix[
      matrix of nodes,
      anchor=north,
      inner sep=0.2em,
      font=\scriptsize
    ]at([xshift=34ex,yshift=-2.2ex]legendpos)
    {
      \ref{plots:plot1}& PR-EoS&[8pt]\\
      \ref{plots:plot2}& Ideal Gas&[8pt]\\
      \ref{plots:plot3}& CoolProp&[4pt]\\};
\end{tikzpicture}
\vspace{-6pt}
\end{center}
\caption{Comparison of PR-EoS and ideal gas EoS with reference data obtained from CoolProp~\cite{CoolProp} for n-hexane at $p = $~5~MPa; a) density, b) specific heat at constant pressure and c) dynamic viscosity.}
\label{fig:validationPR}
\vspace{-6pt}
\end{figure}

At supercritical pressures, intermolecular forces become increasingly important. Using the equation of state (EoS) for an ideal gas would lead to severe deviation from the actual fluid properties, see Fig.~\ref{fig:validationPR}a. Commonly accepted, especially in LES, are cubic EoS and the approximation of real-gas behaviour based on the corresponding state principle. In the present study, the cubic EoS of Peng and Robinson~\cite{peng1976a} (PR-EoS)

\begin{equation}
p = \frac{R \; T}{v - b} - \frac{a}{v^2+2vb-b^2}
\label{eq:PR-EoS}
\end{equation}

is used, where $R$ is the universal gas constant, $T$ is the temperature, $v$ is the molar volume and $a$ and $b$ account for the intermolecular attractive and repulsive forces, respectively. For calculating $a$ and $b$, the concept of a one-fluid mixture in combination with mixing rules is used \cite{poling2001}:

\begin{equation}
a = \sum_{i}^{N_c}\sum_{j}^{N_c} z_i z_j a_{ij} \hspace{1cm} \text{and} \hspace{1cm} b = \sum_{i}^{N_c} z_i b_i \; .
\label{eq:ab_PR-EoS}
\end{equation}

Here, $z_i$ is the mole fraction of species $i$, whereby in the following we denote the overall mole fraction by $\mathbf{z} = \{ z_1 , ... , z_{N_c} \}$ and the liquid and vapor mole fractions by $\mathbf{x} = \{ x_1 , ... , x_{N_c} \}$ and $\mathbf{y} = \{ y_1 , ... , y_{N_c} \}$, respectively. The variables $a_{ij}$ and $b_i$ in Eq.~\eqref{eq:ab_PR-EoS} are calculated based on the corresponding state principle:

\begin{equation}
a_{ij} = 0.45724 \; \frac{R^2 \; T_{c,ij}^2}{p_{c,ij}} \; \left[ 1 + \kappa \left( 1 - \sqrt{\frac{T}{T_{c,ij}}} \right) \right]^2 \hspace{1cm} \text{and} \hspace{1cm} b_i = 0.0778 \; \frac{R \; T_{c,i}}{p_{c,i}} \; ,
\label{eq:aijbi_PR-EoS}
\end{equation}

where $\kappa = 0.37464 + 1.54226 \omega - 0.2699 \omega^2$. The required fluid properties for using the PR-EoS are the critical pressure $p_c$, the critical temperature $T_c$ and the acentric factor $\omega$. For the off-diagonal elements of $a_{ij}$, the pseudo-critical combination rules are used in this study \cite{reid1987}:
\begin{equation}
\begin{array}{l}
\omega_{ij} = 0.5 \left( \omega_i + \omega_j \right) \; , \;\;\; v_{c,ij} = \frac{1}{8} \left( v_{c,i}^{1/3} + v_{c,j}^{1/3} \right)^3 \; , \;\;\; Z_{c,ij} = 0.5 \left( Z_{c,i} + Z_{c,j} \right) \; , \;\;\;\\
\vspace{0pt}\\
T_{c,ij} = \sqrt{T_{c,i} T_{c,j}} \left(1- k_{ij} \right) \hspace{1cm} \textrm{and} \hspace{1cm} p_{c,ij} = Z_{c,ij} R T_{c,ij} /v_{c,ij} \; .\\
\vspace{-8pt}
\end{array}
\label{eq:Harstad}
\end{equation}

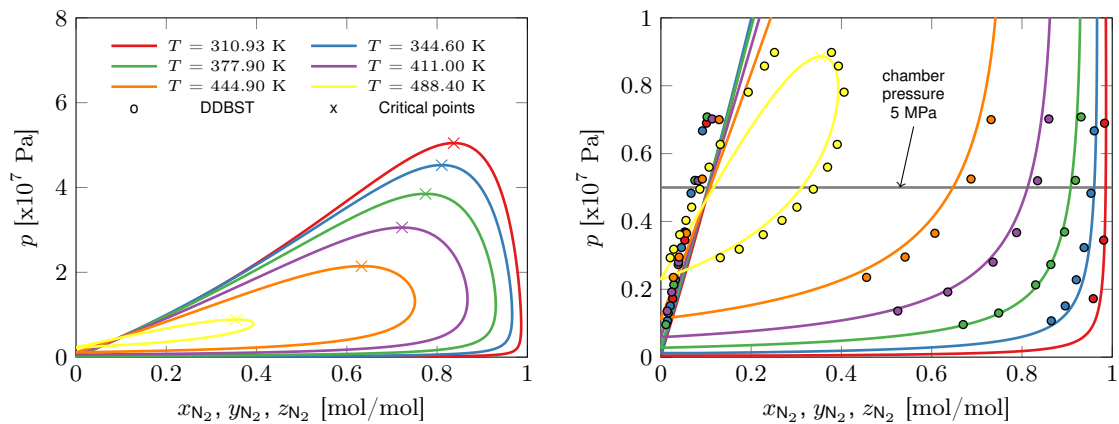
\begin{figure}[h]
\vspace{-12pt}
\begin{center}
\tikzsetnextfilename{vleHexaneNitrogen_kij_0}
\begin{tikzpicture}
\begin{groupplot}[group style = {group size = 3 by 1, horizontal sep = 50pt}, width = 0.47\textwidth, height = 0.38\textwidth]
	\nextgroupplot
	[
	xmin=0,
	xmax=1,
	xlabel = {$x_{\text{N}_{2}}$, $y_{\text{N}_{2}}$, $z_{\text{N}_{2}} \textrm{ [mol/mol]}$},
	x filter/.code={\pgfmathparse{1-#1}\pgfmathresult},
	ymin=0,
	ymax = 8,
	ylabel = $p \textrm{ [x10$^7$ Pa]}$,
	y filter/.code={\pgfmathparse{#1*1e-7}\pgfmathresult}
	]
		\addplot [color=color1,line width=1.0pt] table 	[x=z,y=p]{./data/VLE_Nitrogen_Hexane_310.93K_kij_0.dat};
		\addplot [color=color1,mark=x,mark options={scale=1.5},only	marks] table [y=p_c, x=z_c]{./data/pc_zc_Nitrogen_Hexane_310.93K_kij_0.dat};
		\addplot [color=color2,line width=1.0pt] table 	[x=z,y=p]{./data/VLE_Nitrogen_Hexane_344.6K_kij_0.dat};
		\addplot [color=color2,mark=x,mark options={scale=1.5},only marks] table [y=p_c, x=z_c]{./data/pc_zc_Nitrogen_Hexane_344.6K_kij_0.dat};
		\addplot [color=color3,line width=1.0pt] table 	[x=z,y=p]{./data/VLE_Nitrogen_Hexane_377.9K_kij_0.dat};
		\addplot [color=color3,mark=x,mark options={scale=1.5},only marks] table [y=p_c, x=z_c]{./data/pc_zc_Nitrogen_Hexane_377.9K_kij_0.dat};
		\addplot [color=color4,line width=1.0pt] table 	[x=z,y=p]{./data/VLE_Nitrogen_Hexane_411K_kij_0.dat};
		\addplot [color=color4,mark=x,mark options={scale=1.5},only marks] table [y=p_c, x=z_c]{./data/pc_zc_Nitrogen_Hexane_411K_kij_0.dat};
		\addplot [color=color5,line width=1.0pt] table 	[x=z,y=p]{./data/VLE_Nitrogen_Hexane_444.9K_kij_0.dat};
		\addplot [color=color5,mark=x,mark options={scale=1.5},only marks] table [y=p_c, x=z_c]{./data/pc_zc_Nitrogen_Hexane_444.9K_kij_0.dat};
		\addplot [color=color6,line width=1.0pt] table 	[x=z,y=p]{./data/VLE_Nitrogen_Hexane_488.4K_kij_0.dat};
		\addplot [color=color6,mark=x,mark options={scale=1.5},only	marks] table [y=p_c, x=z_c]{./data/pc_zc_Nitrogen_Hexane_488.4K_kij_0.dat};
		\coordinate (top) at (rel axis cs:0,1);
	\nextgroupplot
	[
	xmin=0,
	xmax=1,
	xlabel = {$x_{\text{N}_{2}}$, $y_{\text{N}_{2}}$, $z_{\text{N}_{2}} \textrm{ [mol/mol]}$},
	x filter/.code={\pgfmathparse{1-#1}\pgfmathresult},
	ymin=0,
	ymax = 1,
	ylabel = $p\textrm{ [x10$^7$ Pa]}$,
	y filter/.code={\pgfmathparse{#1*1e-7}\pgfmathresult}
	]
		\draw [color=gray,line width=1.0pt] (0,0.5) -- (1,0.5);
		\draw [<-] (axis cs:0.53,0.5)--(axis cs:0.56,0.68) node[anchor=south,font=\scriptsize,text width=1.0cm,align=center] {chamber pressure 5 MPa};
		\addplot [color=color1,mark=x,mark options={scale=1.5},only marks] table [y=p_c, x=z_c]{./data/pc_zc_Nitrogen_Hexane_310.93K_kij_0.dat};
		\addplot [color=color1,line width=1.0pt] table [x=z,y=p]{./data/VLE_Nitrogen_Hexane_310.93K_kij_0.dat};\label{plots:VLEplot1}		
		\addplot [color=black,line width=0.5pt,mark=*,mark size=1.5pt,only marks,fill=color1,forget plot] table [x index=2, y expr=\thisrowno{0}*1e5]{./data/DDBST/N2_C6_T_310-93.dat};
		\addplot [color=black,line width=0.5pt,mark=*,mark size=1.5pt,only marks,fill=color1,forget plot] table [x index=4, y expr=\thisrowno{0}*1e5]{./data/DDBST/N2_C6_T_310-93.dat};
		\addplot [color=color2,mark=x,mark options={scale=1.5},only marks] table [y=p_c, x=z_c]{./data/pc_zc_Nitrogen_Hexane_344.6K_kij_0.dat}; 
		\addplot [color=color2,line width=1.0pt] table [x=z,y=p]{./data/VLE_Nitrogen_Hexane_344.6K_kij_0.dat};\label{plots:VLEplot2}		
		\addplot [color=black,line width=0.5pt,mark=*,mark size=1.5pt,only marks,fill=color2,forget plot] table [x index=2, y expr=\thisrowno{0}*1e5]{./data/DDBST/N2_C6_T_344-60.dat};
		\addplot [color=black,line width=0.5pt,mark=*,mark size=1.5pt,only marks,fill=color2,forget plot] table [x index=4, y expr=\thisrowno{0}*1e5]{./data/DDBST/N2_C6_T_344-60.dat};
		\addplot [color=color3,mark=x,mark options={scale=1.5},only marks] table [y=p_c, x=z_c]{./data/pc_zc_Nitrogen_Hexane_377.9K_kij_0.dat};
		\addplot [color=color3,line width=1.0pt] table [x=z,y=p]{./data/VLE_Nitrogen_Hexane_377.9K_kij_0.dat};\label{plots:VLEplot3} 
		\addplot [color=black,line width=0.5pt,mark=*,mark size=1.5pt,only marks,fill=color3,forget plot] table [x index=2, y expr=\thisrowno{0}*1e5]{./data/DDBST/N2_C6_T_377-90.dat};
		\addplot [color=black,line width=0.5pt,mark=*,mark size=1.5pt,only marks,fill=color3,forget plot] table [x index=4, y expr=\thisrowno{0}*1e5]{./data/DDBST/N2_C6_T_377-90.dat};
		\addplot [color=color4,mark=x,mark options={scale=1.5},only marks] table [y=p_c, x=z_c]{./data/pc_zc_Nitrogen_Hexane_411K_kij_0.dat}; 
		\addplot [color=color4,line width=1.0pt] table [x=z,y=p]{./data/VLE_Nitrogen_Hexane_411K_kij_0.dat};\label{plots:VLEplot4}
		\addplot [color=black,line width=0.5pt,mark=*,mark size=1.5pt,only marks,fill=color4,forget plot] table [x index=2, y expr=\thisrowno{0}*1e5]{./data/DDBST/N2_C6_T_411-00.dat};
		\addplot [color=black,line width=0.5pt,mark=*,mark size=1.5pt,only marks,fill=color4,forget plot] table [x index=4, y expr=\thisrowno{0}*1e5]{./data/DDBST/N2_C6_T_411-00.dat};
		\addplot [color=color5,mark=x,mark options={scale=1.5},only marks] table [y=p_c, x=z_c]{./data/pc_zc_Nitrogen_Hexane_444.9K_kij_0.dat}; 
		\addplot [color=color5,line width=1.0pt] table [x=z,y=p]{./data/VLE_Nitrogen_Hexane_444.9K_kij_0.dat};\label{plots:VLEplot5}
		\addplot [color=black,line width=0.5pt,mark=*,mark size=1.5pt,only marks,fill=color5,forget plot] table [x index=2, y expr=\thisrowno{0}*1e5]{./data/DDBST/N2_C6_T_444-90.dat};
		\addplot [color=black,line width=0.5pt,mark=*,mark size=1.5pt,only marks,fill=color5,forget plot] table [x index=4, y expr=\thisrowno{0}*1e5]{./data/DDBST/N2_C6_T_444-90.dat};
		\addplot [color=color6,mark=x,mark options={scale=1.5},only marks] table [y=p_c, x=z_c]{./data/pc_zc_Nitrogen_Hexane_488.4K_kij_0.dat}; 
		\addplot [color=color6,line width=1.0pt] table [x=z,y=p]{./data/VLE_Nitrogen_Hexane_488.4K_kij_0.dat};\label{plots:VLEplot6}
		\addplot [color=black,line width=0.5pt,mark=*,mark size=1.5pt,only marks,fill=color6,forget plot] table [x index=2, y expr=\thisrowno{0}*1e5]{./data/DDBST/N2_C6_T_488-40.dat};
		\addplot [color=black,line width=0.5pt,mark=*,mark size=1.5pt,only marks,fill=color6,forget plot] table [x index=4, y expr=\thisrowno{0}*1e5]{./data/DDBST/N2_C6_T_488-40.dat};
		\coordinate (bot) at (rel axis cs:1,0);
\end{groupplot}	
\path (top|-current bounding box.north)--
      coordinate(legendpos)
      (bot|-current bounding box.north);
\matrix[
      matrix of nodes,
      anchor=north,
      inner sep=0.2em,
      font=\scriptsize
    ]at([xshift=-23.0ex,yshift=-2.2ex]legendpos)
    {
      \ref{plots:VLEplot1}& $T \textrm{ = 310.93 K}$&[5pt]
      \ref{plots:VLEplot2}& $T \textrm{ = 344.60 K}$&[5pt] \\
      \ref{plots:VLEplot3}& $T \textrm{ = 377.90 K}$&[20pt]
      \ref{plots:VLEplot4}& $T \textrm{ = 411.00 K}$&[5pt]\\
      \ref{plots:VLEplot5}& $T \textrm{ = 444.90 K}$&[5pt]
      \ref{plots:VLEplot6}& $T \textrm{ = 488.40 K}$&[20pt] \\
      o & DDBST & x & Critical points\\};
\end{tikzpicture}
\vspace{-12pt}
\end{center}
\caption{Vapour-liquid equilibria at constant temperatures for the investigated binary n-hexane nitrogen mixture; reference data from Dortmund Data Bank Software \& Separation Technology (DDBST), Poston~et~al.~\cite{poston1966a} and Eliosa~et~al.~\cite{eliosa2007a}.}
\label{fig:validationVLE}
\vspace{0pt}
\end{figure}

It has to be noted that in this study the PR-EoS was used in a predictive manner, meaning the binary interaction parameter $k_{ij}$ in Eq.~\eqref{eq:Harstad} was set to zero. The comparison with experimental data of vapor-liquid equilibria, however, shows good agreement for the investigated pressure around 5 MPa, see Fig.~\ref{fig:validationVLE}. Especially the dew-point line shows very good agreement, which is of essential importance in this study.\\

Caloric properties are computed with the departure function formalism, Poling~et~al.~\cite{poling2001}. The reference condition is determined using the nine-coefficient NASA polynomials proposed by Goos~et~al.~\cite{NASA}. The empirical correlation of Chung~et~al.~\cite{chung1988a} is used for the calculation of the viscosity and the thermal conductivity. For pure n-hexane the present thermodynamic model and the reference data from CoolProp~\cite{CoolProp} show excellent agreement, see Fig.~\ref{fig:validationPR}.\\

In this study, the single-phase dense-gas approach was extended with a vapor-liquid equilibrium model which is inspired by the work of Qiu~and~Reitz~\cite{qiu2015a}. In this approach, the single-phase solution of the mixture is considered stable if and only if the Gibbs energy is at its global minimum, see, e.g., Michelsen~and~Mollerup~\cite{michelsen2007}. To check whether a mixture is stable or not, the tangent plane distance (TPD) method of Michelsen~\cite{michelsen1982a}
\begin{equation}
TPD \left( \mathbf{w} \right) = \sum_{i} w_i \left[ \ln w_i  + \ln \varphi_i \left(\mathbf{w}\right) - \ln z_i  - \ln \varphi_i \left(\mathbf{z}\right) \right]
\label{eq:TPD}
\end{equation}
is used, where $\mathbf{w} = \{ w_1 , ... , w_{N_c} \}$ is a trial phase composition and $\varphi_i$ is the fugacity coefficient of component $i$ calculated from the PR-EoS. If the TPD-analysis leads to a negative value for any of the trial phase compositions $\mathbf{w}$, the mixture is unstable and a separation in two phases, namely a vapor~($v$) and a liquid~($l$) phase, is done yielding a decrease in Gibbs energy. It is assumed that this phase split occurs instantaneously and an adiabatic flash is solved. The solution is characterized by the equality of the fugacities $f$ of each component $i$ in the considered phases at a given temperature, pressure and overall composition, i.e. $f_{i,l} (p,T,\mathbf{x}) = f_{i,v} (p,T,\mathbf{y})$. Further details on the implementation can be found in Matheis~and~Hickel~\cite{matheis2016a}. For details on the numerical discretization of the governing equations and turbulence model we refer to Hickel~et~al.~\cite{hickel2014a}.

\section{Test case definition}

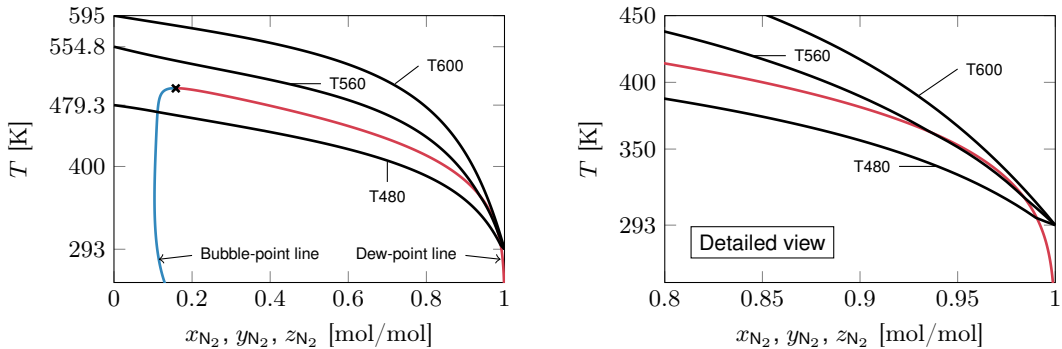
\begin{figure}[h]
	\vspace{-0pt}
	\begin{center}
		\tikzsetnextfilename{adiabaticMixture_kij_0}
		\begin{tikzpicture}
		\begin{groupplot}[group style = {group size = 2 by 1, horizontal sep = 60pt}, width = 0.42\textwidth, height = 0.32\textwidth]
		\nextgroupplot
		[
		xmin=0,
		xmax=1,
		xlabel = {$x_{\text{N}_{2}}$, $y_{\text{N}_{2}}$, $z_{\text{N}_{2}} \textrm{ [mol/mol]}$},
		x filter/.code={\pgfmathparse{1-#1}\pgfmathresult},
		ymin=250,
		ymax = 595,
		ytick={293,400,479.3,554.8,595.0},
		ylabel = $T\textrm{ [K]}$
		]
		\addplot [color=blueVLE,line width=1.0pt] table 	[x=x,y=T]{./data/VLE_xy_Nitrogen_Hexane_5000000Pa_kij_0.dat};\label{plots:adiabaticMixturekij0BP}
		\addplot [color=redVLE,line width=1.0pt] table [x=y,y=T]{./data/VLE_xy_Nitrogen_Hexane_5000000Pa_kij_0.dat};\label{plots:adiabaticMixturekij0DP}
		\addplot [color=black,mark = x,line width=1.0pt,forget plot,only marks] table 	[x=z_c,y=T_c]{./data/Tc_zc_Nitrogen_Hexane_5000000Pa_kij_0.dat};\label{plots:adiabaticMixturekij0CP}
		\addplot [color=black,line width=1.0pt] table [x=z,y=T_EQ]{./data/calcIsenthalpicMixture_F_EQ_Nitrogen-293K-5000000Pa_Hexane-479.3K_Tmix_kij_0.dat};
		\addplot [color=black,line width=1.0pt] table [x=z,y=T_EQ]{./data/calcIsenthalpicMixture_F_EQ_Nitrogen-293K-5000000Pa_Hexane-554.8K_Tmix_kij_0.dat};
		\addplot [color=black,line width=1.0pt] table [x=z,y=T_EQ]{./data/calcIsenthalpicMixture_F_EQ_Nitrogen-293K-5000000Pa_Hexane-595K_Tmix_kij_0.dat};
		\draw (axis cs:0.72,507)--(axis cs:0.78,530) node[anchor=west,font=\scriptsize] {T600} ;
		\draw (axis cs:0.45,507)--(axis cs:0.53,507) node[anchor=west,font=\scriptsize] {T560} ;
		\draw (axis cs:0.7,406)--(axis cs:0.7,380) node[anchor=north,font=\scriptsize] {T480} ;
		\draw [<-] (axis cs:0.115,280)--(axis cs:0.20,285) node[anchor=west,font=\scriptsize] {Bubble-point line} ;
		\draw [<-] (axis cs:0.99,280)--(axis cs:0.90,285) node[anchor=east,font=\scriptsize] {Dew-point line} ;
		\nextgroupplot
		[
		xmin=0.8,
		xmax=1.0,
		xtick={0.8,0.85,0.9,0.95,1.0},
		xlabel = {$x_{\text{N}_{2}}$, $y_{\text{N}_{2}}$, $z_{\text{N}_{2}} \textrm{ [mol/mol]}$},
		x filter/.code={\pgfmathparse{1-#1}\pgfmathresult},
		ymin=250,
		ymax = 450,
		ytick={293,350,400,450},
		ylabel = $T\textrm{ [K]}$,
		tick label style={/pgf/number format/fixed}
		]
		\addplot [color=blueVLE,line width=1.0pt] table [x=x,y=T]{./data/VLE_xy_Nitrogen_Hexane_5000000Pa_kij_0.dat};
		\addplot [color=redVLE,line width=1.0pt] table [x=y,y=T]{./data/VLE_xy_Nitrogen_Hexane_5000000Pa_kij_0.dat};
		\addplot [color=black,mark = x,line width=1.0pt,forget plot] table [x=z_c,y=T_c]{./data/Tc_zc_Nitrogen_Hexane_5000000Pa_kij_0.dat};
		\addplot [color=black,line width=1.0pt] table [x=z,y=T_EQ]{./data/calcIsenthalpicMixture_F_EQ_Nitrogen-293K-5000000Pa_Hexane-479.3K_Tmix_kij_0.dat};
		\addplot [color=black,line width=1.0pt] table [x=z,y=T_EQ]{./data/calcIsenthalpicMixture_F_EQ_Nitrogen-293K-5000000Pa_Hexane-554.8K_Tmix_kij_0.dat};
		\addplot [color=black,line width=1.0pt] table [x=z,y=T_EQ]{./data/calcIsenthalpicMixture_F_EQ_Nitrogen-293K-5000000Pa_Hexane-595K_Tmix_kij_0.dat};
		\draw (axis cs:0.93,390)--(axis cs:0.95,410) node[anchor=west,font=\scriptsize] {T600} ;
		\draw (axis cs:0.845,420)--(axis cs:0.862,420) node[anchor=west,font=\scriptsize] {T560} ;
		\draw (axis cs:0.94,337)--(axis cs:0.92,337) node[anchor=east,font=\scriptsize] {T480} ;
		\draw (axis cs:0.85,268) node[draw,anchor=south,font=\small] {Detailed view} ;
		\end{groupplot}
		\end{tikzpicture}
	\end{center}
	\vspace{-6pt}
	\caption{Adiabatic mixtures of n-hexane and nitrogen at 5 MPa.}
	\label{fig:adiabaticMixture}
	\vspace{-0pt}
\end{figure}

A binary system consisting of n-hexane and nitrogen was chosen and a systematic study was carried out at supercritical pressure with respect to the pure components value ($p_{c,\textrm{C}_6\textrm{H}_{14}}$ = 3.0~MPa and $p_{c,\textrm{N}_2}$ = 3.4~MPa), where n-hexane is injected into a quiescent nitrogen atmosphere at (cold) ambient temperature $T_{ch}$, see Fig.~\ref{fig:exp_setup} and Tab.~\ref{tab:ExperimentalConditions}. The pressure for all test cases was set to $p_{ch}$ = 5 MPa, which correspond to a reduced pressure $p_{r}$ = $p_{ch}/p_{c,\textrm{C}_6\textrm{H}_{14}}$ = 1.67 for n-hexane and $p_{r}$ = $p_{ch}/p_{c,\textrm{N}_2}$ = 1.47 for nitrogen. Three different injection
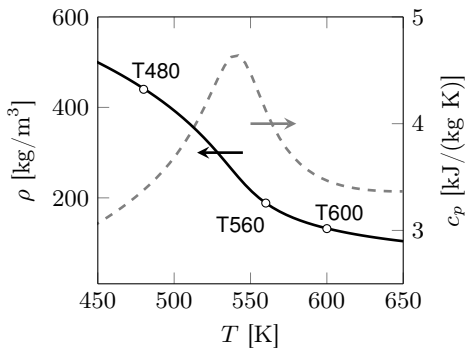
\begin{wrapfigure}{l}{0.45\textwidth}
\vspace{-10pt}
\begin{center}
\tikzsetnextfilename{experimentalConditions}
\begin{tikzpicture}
	\begin{axis}[	
	width=0.35\textwidth,
	height=0.32\textwidth,
	at={(0\textwidth,0\textwidth)},
	xmin=450,
	xmax=650,
	xtick={450,500,550,600,650},
	xlabel = {$T \textrm{ [K]}$},
	axis y line*=right,
	ymin=2.5,
	ymax = 5,
	ylabel = $c_{p} \textrm{ [kJ/(kg K)]}$,
	y filter/.code={\pgfmathparse{#1*1e-3}\pgfmathresult}
	]		
		\addplot [color=gray,line width=1.0pt,dashed] table [x=T,y=cpPR]{./data/val_Hexane_p_5MPa.dat};	
		\draw[>=stealth,->,line width=1.0pt,color=gray](axis cs:550,4)--(580,4);	
	\end{axis}
	\begin{axis}[	
	width=0.35\textwidth,
	height=0.32\textwidth,
	at={(0\textwidth,0\textwidth)},
	axis x line=none,
	xmin=450,
	xmax=650,
	axis y line*=left,
	ymin= 10,
	ymax = 600,
	ylabel = $\rho \textrm{ [kg/m$^3$]}$
	]
		\addplot [color=black,line width=1.0pt] table [x=T,y=rhoPR]{./data/val_Hexane_p_5MPa.dat};
		\draw[fill=white] (axis cs:480,440) circle (1.5pt) node[anchor=south,xshift=1.0ex,yshift=0.2ex,font=\small] {T480};
		\draw[fill=white] (axis cs:560,188.5) circle (1.5pt) node[anchor=north,xshift=-2.0ex,yshift=-0.4ex,font=\small] {T560};
		\draw[fill=white] (axis cs:600,132) circle (1.5pt) node[anchor=south,xshift=1.0ex,font=\small] {T600};
		\draw[>=stealth,->,line width=1.0pt](axis cs:545,300)--(515,300);	
	\end{axis}
\end{tikzpicture}
\end{center}
\vspace{-8pt}
\caption{Density $\rho$ and heat capacity $c_p$ of n-hexane at 5~MPa. Symbols mark injection conditions.}
\vspace{0pt}
\label{fig:ExperimentalConditions}
\end{wrapfigure}
temperatures were selected for the present study as defined in Tab.~\ref{tab:ExperimentalConditions} and visualized in Fig.~\ref{fig:ExperimentalConditions}, based on a-priori calculations of the adiabatic mixture considering two-phase separation; for details see, e.g., Qiu~and~Reitz~\cite{qiu2015a}. Case T480 shows strong two-phase effects, see Fig.~\ref{fig:adiabaticMixture}, and the adiabatic mixture fully penetrates the VLE at 5 MPa. For case T560, only minor two-phase effects can be expected, see Fig.~\ref{fig:adiabaticMixture} detailed view, and the 600 K case is expected to be a dense-gas jet. In addition to the potential phase separation, n-hexane shows strong real-gas effects in the temperature range between 450~K and 650~K, see Fig.~\ref{fig:ExperimentalConditions}, which can be seen in terms of a strong peak in the heat capacity at around 550~K. This is the point where the fluid crosses its Widom-line, see, e.g., Simeoni~et~al.~\cite{simeoni2010a}, and where the fluid undergoes a drastic change from a liquid-like to a gas-like density within a finite temperature interval. The temperatures selected as case names correspond to the total temperature in the injector reservoir $T_{t,\textrm{C}_6\textrm{H}_{14}}$.  The static temperature $T_{\textrm{C}_6\textrm{H}_{14}}$ as well as the outlet velocity at the nozzle exit $u_{\textrm{C}_6\textrm{H}_{14}}$ were calculated based on the assumption of an isentropic nozzle flow \cite{baab2014a} and were used as boundary condition in the LES.

\begin{table}[t]
\vspace{-0pt}
\centering
\caption{Overview of the experimental conditions and boundary conditions used in the LES.}
\vspace{0pt}
\begin{tabular}{c|cccccc}
\toprule
  Case & $p_{ch}$  & $T_{ch}$  & $T_{t,\textrm{C}_6\textrm{H}_{14}}$ & $T_{\textrm{C}_6\textrm{H}_{14}}$ & $u_{\textrm{C}_6\textrm{H}_{14}}$ & $\rho_{\textrm{C}_6\textrm{H}_{14}}$ \\
  & [MPa] & [K] & [K] & [K] & [m/s]  & [kg/m$^3$]\\
\midrule 
T480 & 5.0 & 293.0 & 480.0 & 479.3 & 51.0 & 443.2\\
T560 & 5.0 & 293.0 & 560.0 & 554.8 & 72.1 & 202.0\\
T600 & 5.0 & 293.0 & 600.0 & 595.0 & 90.3 & 136.9\\
\end{tabular}
\label{tab:ExperimentalConditions}
\vspace{-12pt}
\end{table}

\section{Results and discussion}

\begin{figure}[p]
\begin{center}
\begin{tabular}{x{11.5cm} x{0.3cm}}
\includegraphics[width = 0.72\textwidth]{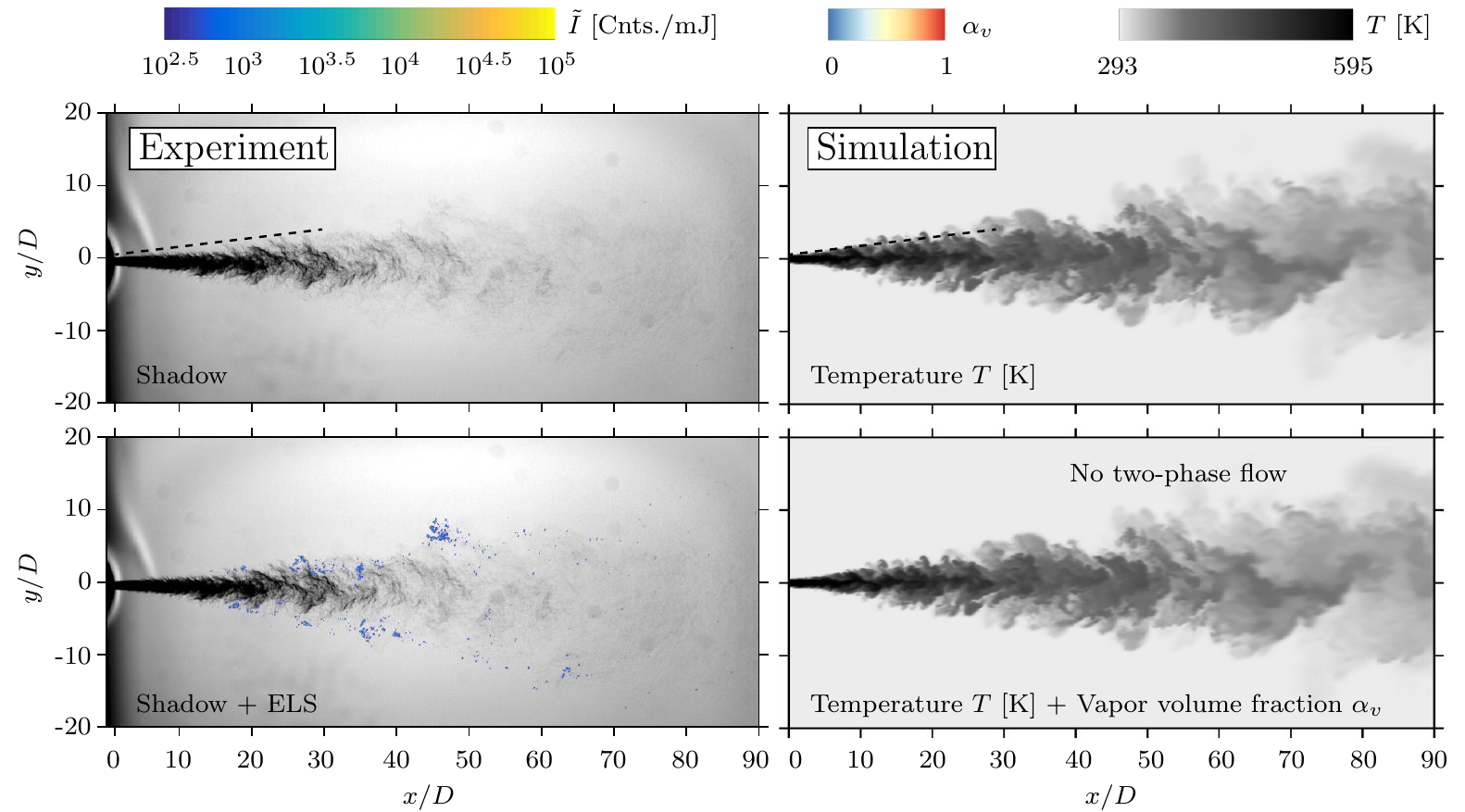} & \rotatebox{90}{\hspace{-120pt}a) Case T600}\\
\includegraphics[width = 0.72\textwidth]{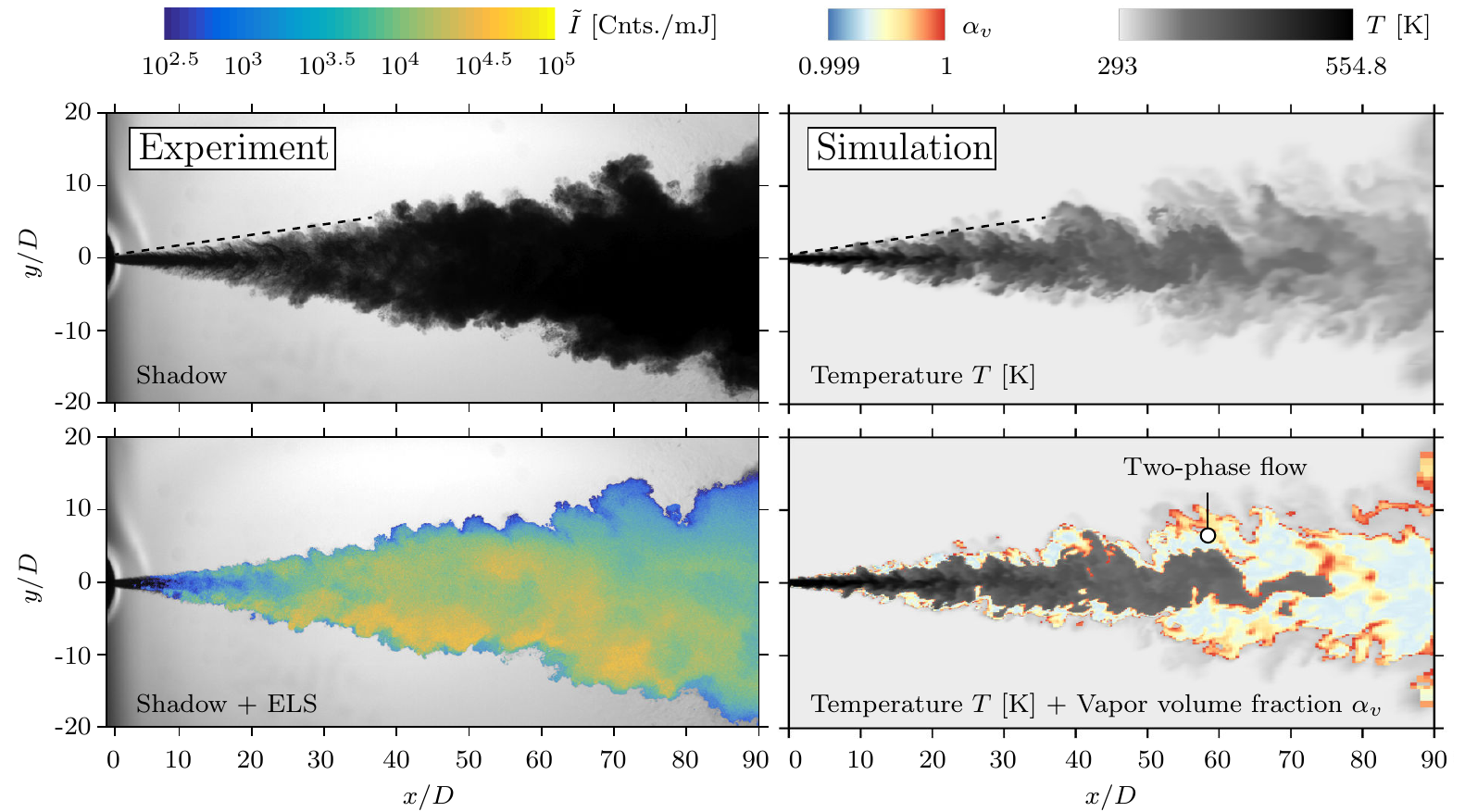} & \rotatebox{90}{\hspace{-120pt}b) Case T560}\\
\includegraphics[width = 0.72\textwidth]{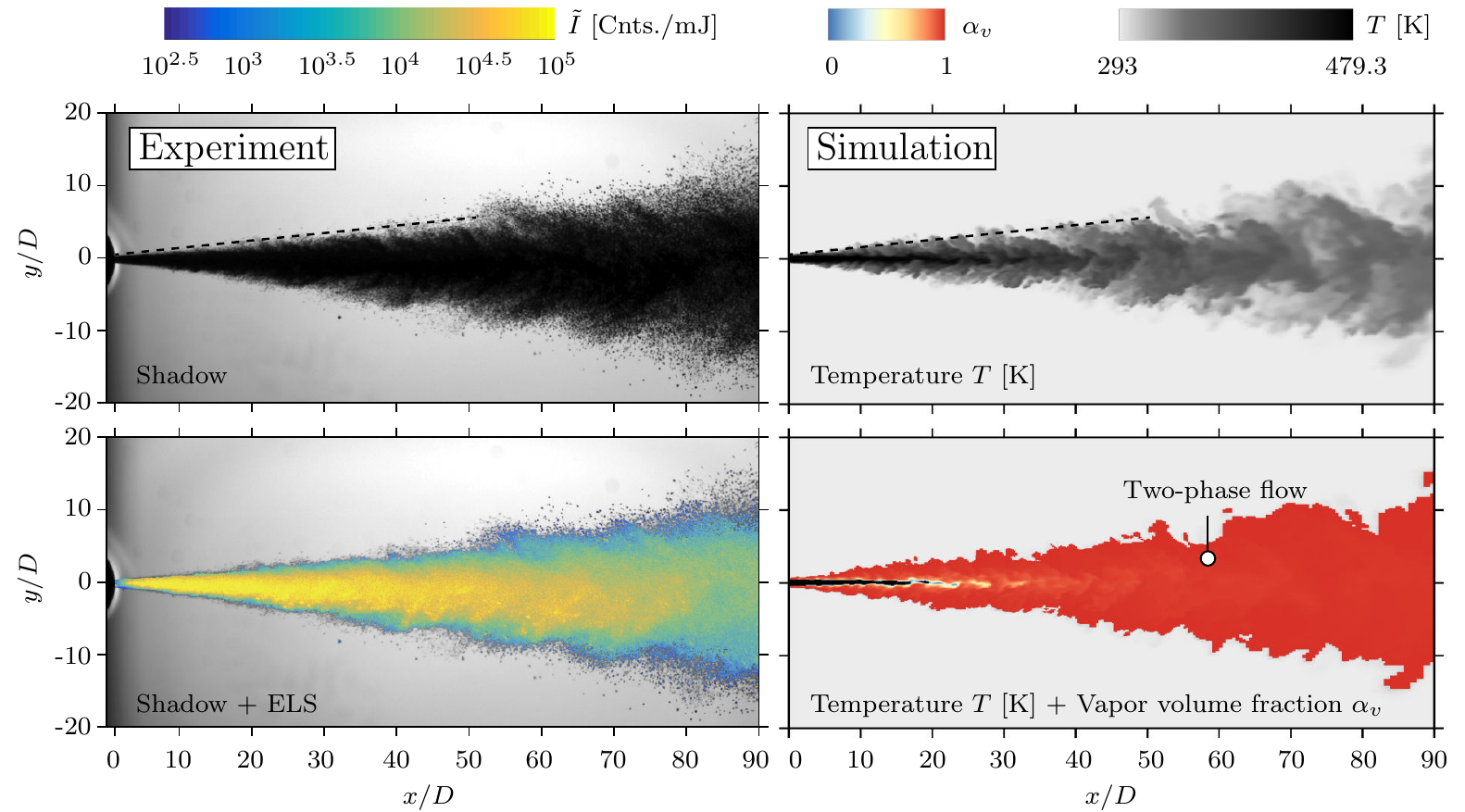} & \rotatebox{90}{\hspace{-120pt}c) Case T480}
\end{tabular}
\end{center}
\vspace{-12pt}
\caption{Comparison of experimental and numerical snapshots for the cases T600, T560 and T480.}
\label{fig:comparisonExpSim}
\end{figure}

In Fig.~\ref{fig:comparisonExpSim}, experimental single-shot measurements and instantaneous LES results are compared. Snapshots were taken at a time sufficiently large such that the jets are fully developed and can be considered as quasi-stationary. Shadowgram (top frame) and simultaneously taken ELS image superimposed onto the corresponding shadowgram (bottom frame) visualize both the flow structure as well as the thermodynamic state. The intensity $\tilde{I}$ is shown on a logarithmic scale comprising almost three orders of magnitude. Focus shall hence be put on the overall scattering characteristics rather than quantitative evaluation of the measured signals. Values outside the color scale range were cut off to emphasize the distinction between regions of negligible $\tilde{I}$ and two-phase regions within the jet, indicated by high scattering intensities. In the LES, flow structures are visualized by the instantaneous temperature field (top frame). In the bottom frame the vapor volume fraction $\alpha_v$ is superimposed to indicate regions of two-phase flow. By doing so, a direct comparison of the phase formation phenomena in the experiment and LES is provided.

\begin{figure}[h]
\vspace{0pt}
\begin{center}
\begin{overpic}[width=0.325\textwidth]{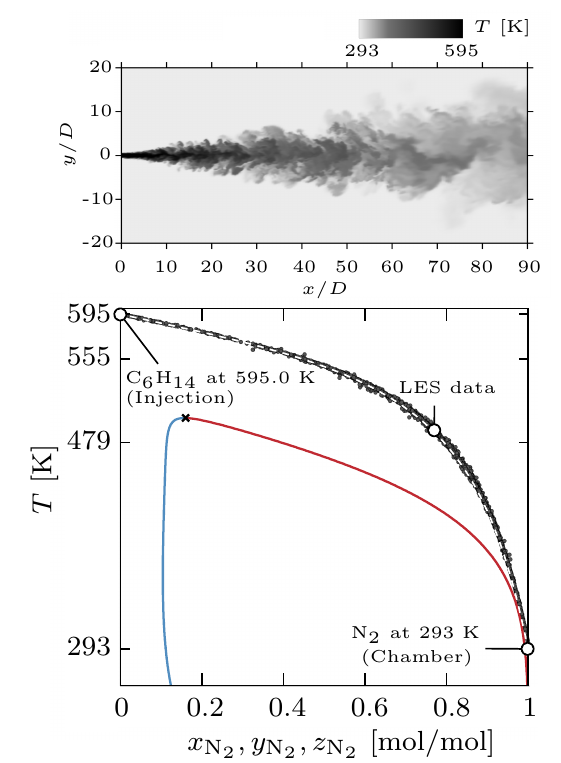}%
  \put(3,94){\small a)}%
\end{overpic}
\begin{overpic}[width=0.325\textwidth]{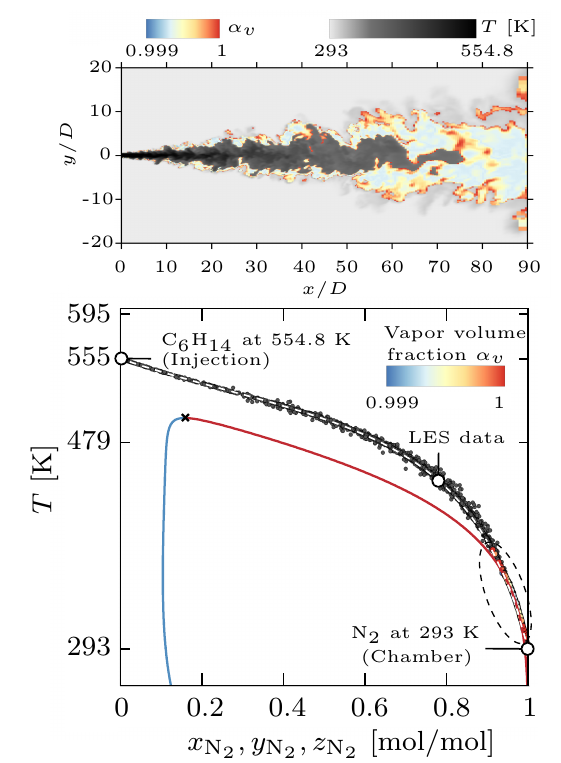}%
  \put(3,94){\small b)}%
\end{overpic}
\begin{overpic}[width=0.325\textwidth]{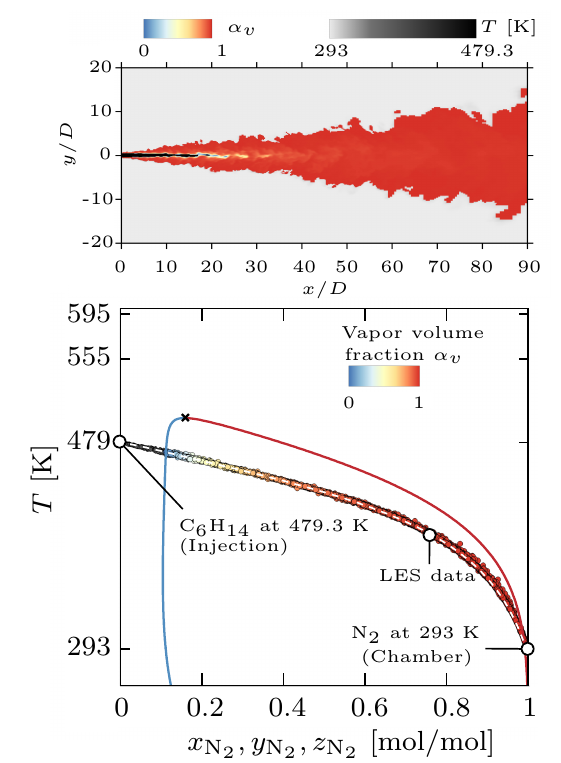}%
  \put(3,94){\small c)}%
\end{overpic}
\end{center}
\vspace{-6pt}
\caption{Scatter plots of the cases T600, T560 and T480 showing the thermodynamic state of the LES results together with the vapor-liquid equilibrium at 5 MPa and the adiabatic mixture.}
\label{fig:scatterPlots}
\vspace{6pt}
\end{figure}

The experimental results for Case T600 show a dark core being dissolved in the environment and finger-like structures emerging from the surface of the dark core, see Fig.~\ref{fig:comparisonExpSim}a. No significant (stable) scattering signal is measured and as a result a single-phase state can be deduced. This finding agrees well with both the LES results and the adiabatic mixing line. The thermodynamic model does not predict any thermodynamically unstable states based on the applied tangent plane distance criterion, see Eq.~\eqref{eq:TPD}. The stability of the single phase in the LES can be seen in more detail in Fig.~\ref{fig:scatterPlots}a, where all conditions are presented by means of a scatter plot. All conditions in the LES follow the adiabatic mixture line closely and none of the points lies within the region of the vapor-liquid equilibrium at 5~MPa. It is worth mentioning that the jets being investigated are almost isobaric and are therefore not subjected to large fluctuations in pressure. Furthermore, the dew-point line does not show a strong dependency on the pressure close to the investigated pressure of 5~MPa, see Fig.~\ref{fig:validationVLE}, underlining the experimental observation that a pronounced two-phase region within the jet does not exist. The minor signals determined in the experiment might stem from spatial fluctuations of thermodynamic properties, which may lead to small areas of phase separation. Downstream of $x/D \gtrsim 10$, these regions form in the outer shear layer but they are highly unstable and collapse instantaneously. Another reason for the detected scattering signal might be a local violation of the adiabatic mixture assumption due to, for instance, heat conduction.

A totally different picture results for case T560, see Fig.~\ref{fig:comparisonExpSim}b. While the shadowgram resembles that for T600 in the near-nozzle region, a strong difference becomes apparent for $x/D \gtrsim 20$. Here, a dark region forms over the entire extension of the jet, which indicates the formation of a dense droplet cloud. This fact is clearly corroborated by a steep increase in the scattering intensity, which is initiated by droplet generation in the mixing layer for $x/D \gtrsim 10$. Further downstream, this flow mixes into the jet core, leading to a pronounced two-phase characteristic throughout the entire jet domain. In fact, this is proved by the LES result of case T560, where a large spatial area of the jet shows two-phase behaviour due to an unstable single-phase mixture determined by the TPD-criterion, see Eq.~\eqref{eq:TPD}. In agreement with the experimental observation, the LES regions showing two-phase flow are located in the outer shear layer, which merge towards the jet centerline downstream of $x/D \gtrsim 60$. Due to the only minor penetration of the VLE, the vapor volume fraction in the regions with phase separation is very close to unity, see Fig.~\ref{fig:scatterPlots}b. In the LES, the phase separation is occurring in the shear layer first where all mixture states are present, forming a layer of two-phase flow around the actual jet. Further downstream, this layer is growing steadily and, as additional nitrogen mixes with the jet, this leads to a progressing dilution and to a cooling of the mixture, see Fig.~\ref{fig:comparisonExpSim}b. For $x/D \gtrsim 60$, this steady mixing and cooling results in a jet where a huge portion exhibits two-phase behaviour, which becomes obvious by comparing the LES snapshot and the $T$,$z$-diagram in Fig.~\ref{fig:scatterPlots}b. As stated above, this is in very good agreement with the experimental results. The LES vapor-liquid equilibrium model is able to phenomenologically capture the transition from a dense-gas mixture (case T600) to a jet exhibiting phase separation (case T560). In addition, identical to the case T600, all states in the LES group around the adiabatic mixture line as predicted by the a-priori analysis, see Fig.~\ref{fig:scatterPlots}b. 

\afterpage{\clearpage}

For Case T480, a drastic change in terms of flow phenomena occurs. A spray-like behaviour similar to atomized jets \cite{chehroudi1999a,mayer1998a} can be seen in Fig.~\ref{fig:comparisonExpSim}c. The corresponding shadowgram shows a constant dark core with distinct droplets visible in the outer jet region for $x/D \gtrsim 15$. This pronounced two-phase characteristic directly reflects in the strong scattering throughout virtually the entire jet domain. As in case T560, the presence of this strong phase separation is confirmed by the LES where nearly the complete jet has entered the VLE, see Fig.~\ref{fig:scatterPlots}c. In contrast to the cases T560 and T600, the adiabatic mixture of case T480 crosses both the bubble-point line as well as the dew-point line and the corresponding shadowgram shows a jet exhibiting spray-like behaviour. Due to the Eulerian approach used in the LES, no individual droplets are resolved and, therefore, the spray-like character is represented in terms of the vapor fraction covering the whole range from zero to one. A single-phase state is predicted only in the ``dark'' core at the injector inlet and in a thin area at the outermost region of the shear layer, see Figs.~\ref{fig:comparisonExpSim}c~and~\ref{fig:scatterPlots}c. As soon as the mixture crosses the bubble-point line at $z_{\textrm{N}_2} \approx 0.1$, the jet exhibits two-phase behaviour. Due to the approximate parallelism of the adiabatic mixture line and the dew-point line, high vapor volume fractions larger than 0.5 result from $z_{\textrm{N}_2} \approx 0.3$ on and cause the reddish color in the contour-plot of the jet in Figs.~\ref{fig:comparisonExpSim}c~and~\ref{fig:scatterPlots}c.

Apart from the good prediction of the LES with respect to the thermodynamic effects, the opening angle of the jet shows a generally good agreement between experiments and numerical simulations for all three cases. For case T480, a smaller opening angle compared to the cases T560 and T600 is observed. For case T600, the LES appears to overestimate the opening angle slightly. It has to be noted, however, that shadowgraphy may be too insensitive to capture the entire jet domain as given by the LES.

\section{Conclusions}

Experiments and numerical simulations of jet mixing at high pressure were carried out at three different research facilities in Germany in order to contribute to a better understanding and prediction of high-pressure injection and the phase separation processes in initially supercritical fluids. A multi-component system consisting of n-hexane and nitrogen was chosen and a systematic study was conducted at supercritical pressure with respect to the pure components value. N-hexane was injected into a chamber filled with pure nitrogen at (cold) ambient temperature. The test case conditions were selected such that they cover regimes in which the thermodynamic non-idealities, in particular the effects that stem from the potential phase separation, are significant. Three different test cases have been presented and discussed in this paper. Simultaneous shadowgraphy and elastic light scattering experiments were conducted in order to capture both the flow structure as well as the phase separation. In addition, numerical simulations were carried out by means of large-eddy simulations with a vapor-liquid equilibrium model. Experimental and numerical results show phase separation and the transition from a dense-gas to a spray-like jet, where the a-priori calculation predicts two-phase flow. Characteristics of the formation process agree well between experiments and numerical simulations. The formation of a two-phase flow is initiated in the mixing layer some distance downstream of the nozzle and eventually mixes into the jet core at large distance.

This study serves as a basis for more thorough investigation of these kind of jets. A more detailed examination of the near injector region as well as a comparison of transient and averaged data from experiments and numerical simulations will be conducted.

\section*{Acknowledgements}

The authors gratefully acknowledge the German Research Foundation (Deutsche Forschungsgemeinschaft) for providing financial
support in the framework of SFB/TRR 40. Financial support was also provided by Munich Aerospace (\url{www.munich-aerospace.de}). Furthermore, the authors thank the Gauss Centre for Supercomputing e.V. (GCS) (\url{www.gauss-centre.eu}) for supporting this project by providing computing time on the GCS Supercomputer SuperMUC at Leibniz Supercomputing Centre (\url{www.lrz.de}).

\bibliography{Bib_ILASS_2017} 
\bibliographystyle{ASME}

\end{document}